\documentclass[10pt,aps,pre,twocolumn,showkeys,floatfix,longbibliography]{revtex4-1}
\usepackage{graphicx}
\usepackage{hyperref}
\usepackage{amsmath,amssymb,amsfonts}

\begin{document}
\title{Pattern formation in desiccated sessile colloidal droplets with salt admixture: Short review}
\author{Yuri Yu. Tarasevich}
\email[Corresponding author: ]{tarasevich@asu.edu.ru}
\affiliation{Astrakhan State University, Astrakhan, Russia}
\author{Sujata Tarafdar}
\email{sujata\_tarafdar@hotmail.com}
\affiliation{Physics Department, Jadavpur University, Kolkata 700032, India}
\author{Tapati Dutta}
\email{tapati_mithu@yahoo.com}
\affiliation{St. Xavier's College, Kolkata 700016, India}
\date{\today}

\begin{abstract}
This short review is devoted to the simple process of drying a multi-component droplet consisting of a complex fluid containing a salt. These processes provide a fascinating subject for study. The explanation of the rich variety of patterns formed is not only an academic challenge, but a problem of practical importance, as applications are growing in medical diagnosis and improvement of coating/printing technology.
The fundamental scientific problem is the study of the mechanism of micro- and
nanoparticle self-organization in open systems. The specific fundamental problem to
be solved, related to this  system is - the investigation of the mass transfer processes, the formation and evolution of phase fronts and the identification of mechanisms of pattern formation.  The  drops of liquid
containing dissolved substances and suspended particles are assumed to be drying on a horizontal solid substrate. The chemical
composition and macroscopic properties of the complex fluid, the concentration and nature of the salt,
the surface energy of the substrate and the interaction between the fluid and substrate which determines the wetting, all affect the final morphology of the dried film.
\end{abstract}

\maketitle

\section{Introduction}\label{sec:intro}

The structures observed after drying  biological fluids on a horizontal impenetrable substrate attracted the attention of European researchers as early as the 1950s~\cite{Koch1954,Koch1956,Koch1957,Koch1957a,Koch1957b,Koch1957c,Sole1954,Sole1955,Sole1957a,Sole1957b,SoleBook}.
Unfortunately, this series of articles did not attract attention from the physics community. In the 1980s the phenomenon has been reopened in the USSR~\cite{Rapis1988eng}. In the 1980s-1990s,  doctors of the Soviet Union and the countries of the former Soviet Union began to use the appearance of structures formed by drying droplets of biological fluids for the diagnosis of various diseases. Numerous articles are published in Russian, in medical journals, dissertations are defended, and different methods are patented that are devoted to the diagnosis of diseases on the basis of the structures formed by drying droplets of biological fluids. The relevant references can be, for example, found in the books~\cite{Savina1999,Shabalin2001,Rapis2002,Vorobev}. Unfortunately, the world scientific community is poorly informed about these works of Soviet and Russian scientists, since only a very few of these articles have been published in English~\cite{Shabalin1996,Yakhno2003,Shatokhina2004,Shabalin2007,Martusevich2007BEBM,Yakhno2015SRP}.

In the last two decades, interest in the structures arising on drying complex fluids has increased  worldwide due to a variety of applications, such as  bio-preservation~\cite{Ragoonanan2008,Less2013},  high-throughput drug screening~\cite{Takhistov2002}, fast identification of fluid and substrate chemistry based on automatic pattern recognition of stains~\cite{Kim2012AM}, assessment of  quality of products~\cite{Kokornaczyk2011}, and Raman spectroscopy~\cite{Esmonde2008AS,Esmonde2008ProcSPIE,Esmonde2009,Esmonde2009PSIE,Filik2007,Filik2008,FilikJRS2009,Zhang2010AS,Pearce2000OPO,Dingari2012,Kocisova2012}. During the high-throughput drug screening, pattern formation in the the drying sample is not desirable~\cite{Takhistov2002}.  Several reviews~\cite{Routh2013,Sefiane2013,Larson2014,Zhong2015,Sadek2015DST,Chen2016ACIS} and books~\cite{Lin2010,WaterDroplets2013,DropletWetting2015,goehring2015desiccation} based on pattern formation during desiccation, published within the short span of just four years confirms the intensely growing interest in this topic.

While performing the experiments seems very simple (table top experiments producing interesting patterns can be done even at home),  understanding the physics behind the  pattern formation  phenomena turns out to be extremely complicated and involves a number of interrelated processes of different nature~\cite{Tarasevich2004,Yakhno2009JTP}.
During desiccation of biological fluids, a  sequence of various physical and physico-chemical processes can be observed~\cite{Yakhno2008JCIS,Yakhno2011Biophys}. For example, redistribution of the components occurs. Protein molecules are carried out by flows to the edge of the droplet, and accumulate to form a gel. The salt is distributed over the whole area of the droplet  almost uniformly. After complete drying of the droplet, a protein precipitate  remains on the substrate in the form of a ring, the width of  the ring depends on concentrations of the protein and the salt~\cite{Shabalin2001,Prokhorov}. Salt crystals can form fractal (dendritic) structures~\cite{Annarelli2001,Tarasevich2003JTP,Gorr2013CSB,Choudhury2013,DuttaChoudhury2015SM}. In the later stages of drying, a sample may crack~\cite{Pauchard1999,Annarelli2001,goehring2015desiccation}, the characteristic pattern of the cracks also helps in diagnosing diseases which the subject may be suffering from~\cite{Yakhno2005c}.

Analysis based on a visual comparison of the structures formed by drying a liquid drop~\cite{Rapis1988eng,Savina1999,Shabalin2001,Rapis2002,Yakhno2003} has  significant drawbacks. Conclusions are liable to be subjective, without techniques for defining quantitative parameters to characterize the structures.
Computer pattern recognition may be tried to eliminate this shortcoming~\cite{Buzoverya2014JTPen,BouZeid2013,Kim2012AM}.

Although pattern formation can be observed during  drying of both inorganic and organic colloids, the case of biological fluids is attracting increasingly growing interest from the scientific community in recent years~\cite{Sobac2011PRE,Brutin2011JFM,Sefiane2010JBE}.

Despite the application of the phenomenon, for practical purposes and considerable progress in the understanding of the phenomenon~\cite{Tarasevich2004,Yakhno2009JTP}, the theoretical description of the pattern formation in desiccating  biological fluids is still incomplete. The physical, biophysical, biochemical, biological, and physico-chemical processes occurring in the dehydration of biological fluids remain largely to be clarified. This explanation is a necessary background for understanding the connections between physico-chemical properties of the biological fluids and observed patterns.

Even though numerous publications in medical literature have been devoted to pattern formation in biological fluids during their drying, physicists, chemists and mathematicians began to pay attention to this problem only recently. During the last decade, several groups of physicists began to actively publish work in this direction~\cite{Bardakov2010FD,Chashechkin2010DPh,Kistovich2010,Brutin2011JFM}. A significant part of the publications from Russia belongs to only two research groups, namely a team from Astrakhan State University~\cite{Tarasevich2007JTP,Tarasevich2005,Tarasevich2010JTP,Tarasevich2009,Tarasevich2004,Tarasevich2007epje,Vodolazskaya2010,Tarasevich2011CPS,Tarasevich2013CSA} and a team from Insitute of Applied Physics RAS~\cite{Yakhno2004JTP,Yakhno2009JTP,Yakhno2005,Yakhno2003,Yakhno2008JCIS,YakhnoNM,Yakhno2005c,Yakhno2015}.
At the same time,  interest in pattern formation during the drying of biological fluids in the community of Physicists is growing rapidly outside Russia. Recently,  a number of works have been published~\cite{BouZeid2013,BouZeid2013COLSUA1,Brutin2012JHT,Brutin2011JFM,Gorr2013CSB,Pradhan2012}.
Some researches are dedicated to the similar systems, namely organic~\cite{Dutta2013COLSUA,Choudhury2013,DuttaChoudhury2015SM,Roy2015ASS}  and inorganic~\cite{Pauchard1999} colloids with salt admixtures.
A considerable part of publications on desiccation of organic colloids with salt admixtures belongs to the Indian researchers~\cite{Basu2012,Choudhury2013,DuttaChoudhury2015SM,Dutta2013COLSUA,Giri2013,Roy2015ASS}.

Most of the studies consider the following scenario: a droplet of liquid dries on a horizontal hydrophilic substrate  with a constant base. Here the fluid-substrate-vapour triple line is pinned and the contact angle decreases~\cite{Parisse1996,Deegan1997,Deegan2000}. Evaporation in the presence of pinning leads to an outward flow within the droplet; this flow carries solute and suspended particles to the edge of the droplet causing  a deposit rim to form at the droplet edge.

The processes occurring during the desiccation of the sessile colloidal droplets and morphology of the resulting precipitate depend on many different factors, e.g. the nature and shape of the colloidal particles~\cite{Yunker2011} and their initial volume fraction~\cite{Yakhno2015}, the presence of admixtures (e.g. surfactants) in the solution~\cite{YakhnoNM,Yakhno2010TP,Still2012,Anyfantakis2015}, ionic strength and pH of the solution~\cite{Pauchard1999}, the properties of substrate (thermal conductivity, whether hydrophilic/hydrophobic)~\cite{Ristenpart2007PRL,Brutin2012JHT,Choudhury2013,Carle2013lang}, evaporation mode~\cite{Caddock2002,Chhasatia2010APL,BouZeid2014CSA,BouZeid2013,BouZeid2013COLSUA1}, etc.

Classification of possible desiccation modes  may be done using two characteristic times, namely, the drying time, $t_\text{d}$, and the gelation time, $t_\text{g}$,~\cite{Pauchard1999}. There are three different modes of colloidal sessile droplet desiccation~\cite{Pauchard1999}:
\begin{enumerate}
 \item $t_\text{g}\gg t_\text{d}$, where $t_\text{g}$ is the gelation time and $t_\text{d}$ is the desiccation time. The gelled phase occurs near the droplet edge and moves inward while the central area of the droplet remains liquid.
 \item $t_\text{g}\approx t_\text{d}$. The gelled skin covers the free droplet surfaces. This thin shell cannot prevent evaporation of the solvent. The buckling instability occurs~\cite{Pauchard2003CRP}.
   \item $t_\text{g}\ll t_\text{d}$. The phase transition from sol to gel in the whole bulk of the droplet is almost instantaneous. The gelled droplet loses solvent via evaporation very slowly.
\end{enumerate}

When $t_\text{g}\gg t_\text{d}$,
 the desiccation process can be divided into several stages (see, \emph{e.g.},~\cite{Okuzono2009,Yakhno2008JCIS,Jung2009}).
\begin{enumerate}
 \item Initial single-phase liquid stage. The whole droplet is a sol. The outward flow carries suspended particles to the droplet edge until the volume fraction of the suspended particles, $\Phi$, reaches the critical value, $\Phi_\text{g}$. Note that particle-\-enriched region is extremely narrow, whereas the particle volume fraction in the central area of the droplet is almost constant along its radius.  This stage was simulated in~\cite{Tarasevich2007JTP,Tarasevich2007epje,Tarasevich2010JTP}, as well as in~\cite{Okuzono2009}.
 \item Intermediate two-phase stage. A Gelled ring appears near the droplet edge  and grows towards the droplet center. The volume fraction of the colloidal particles is constant inside the  \textit{foot} i.e. the outer gelled band, $\Phi_\text{g}$, and almost constant in the sol, $\Phi$, except for a rather narrow area near the phase front.  This stage was simulated in  works~\cite{Vodolazskaya2010,Tarasevich2011CPS,Vodolazskaya2011MPLB}, as well as in~\cite{Ozawa2005,Okuzono2009}.
 \item Final single-phase solid stage. The gelled deposit loses the remaining  moisture very slowly. Some real fluids of interest (\emph{e.g.}, biological fluids) can contain both suspended particles and dissolved substances. In this case, the dendritic crystals can occur in the central area of a sample~\cite{Dutta2013COLSUA,Takhistov2002,Annarelli2001}.  Finally, the desiccation crack patterns appear~\cite{Pauchard1999,Annarelli2001,Leung2001Nature,Neda2002,Golbraikh2003,BouZeid2013,Sobac2014CSA,goehring2015desiccation}.
\end{enumerate}

\begin{figure}
  \centering
  \includegraphics[width=\linewidth]{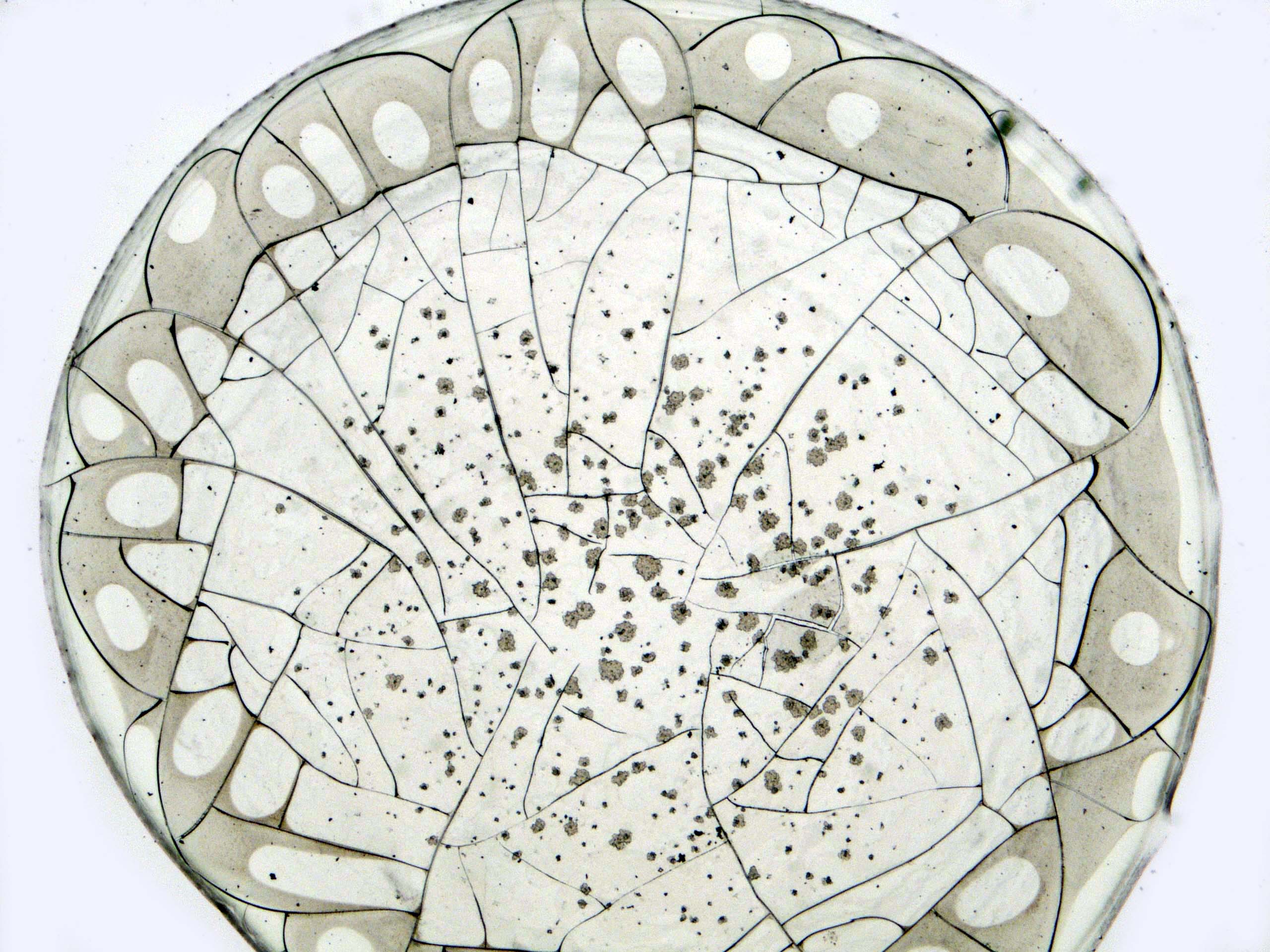}
  \caption{Dried sample of HSA droplet. \href{http://koltovoi.nethouse.ru/}{Courtesy N.A.~Koltovoi} }\label{fig:koltovoi1}
\end{figure}

\section{2D models of mass transfer}\label{sec:model2d}
Modelling of the processes occurring during the drying of colloidal droplet solutions is very complicated, because these processes are extraordinary varied and complex~\cite{Yakhno2008JCIS}. The  authors have different views about the driving mechanisms that lead to the formation of the solid phase~\cite{Deegan2000a,Deegan1997,Deegan2000,Fischer2002,Parisse1996,Popov2005}. For example, \cite{Widjaja2008} considered competition of convection and sedimentation, but~\cite{Okuzono2009} considered competition of convection and diffusion.
Numerous models were proposed during last two decades. Several models describe some particular processes occurring during the colloidal droplet desiccation, e.g. capillary flow, mass transport processes, etc.~\cite{Parisse1996,Deegan2000,Fischer2002,Popov2005,Ozawa2005,Widjaja2008,Bhardwaj2009,Zheng2009,Witten2009,Sefiane2009Langmuir,Petsi2010,Kistovich2010,Kim2011}. Generally, models are developed for systems with low concentrations of the colloidal particles.

Two very different situations are possible when a colloidal  sessile droplet desiccates. In the first case, the particles inside a droplet can interact with each other only mechanically (impacts). In this case, the deposit forms a porous medium. Such a medium prevents neither bulk flow inside it nor evaporation from its surface. Moreover, such a porous medium can enhance evaporation from its surface due to drainage effect~\cite{Bhardwaj2009}.
In the second case, the colloidal particles can form strong inter-particle bonds. In this case, hydrodynamic flows, particle diffusion and solvent evaporation are restricted.
The proposed theoretical models mainly deal with the first situation~\cite{Deegan2000,Bhardwaj2009,Sefiane2009Langmuir,Fischer2002,Kim2011,Kistovich2010,Ozawa2005,Parisse1996,Petsi2010,Popov2005,Widjaja2008,Witten2009,Zheng2009}. Only a few models treat the deposit as impenetrable for flows and preventing evaporation~\cite{Ozawa2005,Okuzono2009,Tarasevich2011CPS,Vodolazskaya2011MPLB}. Nevertheless, the simulation of desiccated colloidal droplets with phase transition is extremely important for high-throughput drug screening~\cite{Takhistov2002}, bio-stabilization~\cite{Ragoonanan2008}, identification of fluids~\cite{Kim2012AM}, and medical tests~\cite{Shabalin1996,Yakhno2005,Killeen2006,Martusevich2007}. The models~\cite{Ozawa2005,Okuzono2009,Tarasevich2011CPS,Vodolazskaya2011MPLB} utilize sets of rather complicated partial differential equations (PDE).

Several models describing desiccated sessile colloidal droplets have been reported recently~\cite{Ozawa2005,Okuzono2009,Tarasevich2011CPS,Eales2015,Eales2015JCIS}. They are based on lubrication approximation~\cite{Anderson1995}. This approach has several serious shortcomings~\cite{Lebovka2014CSA} as enumerated below.
\begin{enumerate}
  \item Only thin films can be considered, all quantities are supposed to be dependent only on one radial coordinate.
  \item In fact, a two phase system is considered as  one-phase, the gel is assumed to be a liquid with very high viscosity, the hydrodynamic equations are written for the whole droplet desiccation.
  \item The mathematical expression for evaporation flux above the free surface is  speculative rather than supported by experiments. To our best knowledge, measurements of the vapour flux above a system with sol-gel phase transition are not published yet.
  \item Knowledge of the effect of particle concentration on viscosity is needed for calculations. This dependence can be obtained from experiments with rather large volumes of colloid. Viscosity of a small droplet with a large free surface and large contact area with a substrate can deviate from this in a rather complex manner.
  \item It is assumed that all the molecules that get to the edge of the droplet pass into the solid phase. Generally, this assumption can be wrong in the presence of convection of any nature in a droplet. An inward flux of particles due to diffusion may also exist.
\end{enumerate}
To overcome the limitations of the listed models, a 3-dimensional (3D) model should be developed and utilized.

\section{Modelling flow in 3-dimensions}\label{sec:model3d}
A number of  papers devoted to 3D  models of processes inside evaporating droplets were published during the few last years. Mostly, the articles considered droplets of pure liquids and simulated flows within them~\cite{Hu2002,Mollaret2004,Tarasevich2005,Widjaja2008CCE,Dunn2008,Barash2009a,Barash2009b}.
The analytical solutions of the Laplace equation, that describe the velocity field inside evaporating droplets of a non-viscous liquid were obtained for the contact angle of 90$^\circ$ by Tarasevish~\cite{Tarasevich2005} and for a case of arbitrary contact angle by Masoud~\cite{Masoud}.
Flow inside the boundary line of an evaporating liquid for any
contact angle were found using Stokes approach~\cite{Petsi2008}. Numerical calculations of the velocity field within evaporating droplets were performed using Finite Element Method~\cite{Larson2005a,Larson2005b,Mollaret2004}. Presence of dissolved substances or suspended particles  inside the droplets and deposit formation were not taken into account  in these models.

\section{3D models of mass transfer}\label{sec:transfer}
3D models describing the processes inside the particle-laden droplets were developed using both the continuum and discrete approaches. Development of discrete models was initiated by the requirements of modelling of evaporation-driven self-assembly (EDSA) or  evaporation-induced self-assembly (EISA)~\cite{Hsu2007PSSA,Kim2011,Chen2013,Crivoi2013CSA,LebedevStepanov2013CSA,Lebovka2014PRE,Crivoi2014,Fujita2015,Hwang2015NHTB}. Additional references can be found in the review~\cite{Fujita2010}. Recently published models considered the Brownian motion of particles inside the droplets. For instance, in the work of Petsi~\cite {Petsi2010} the Brownian motion of the particles is superimposed on the hydrodynamic flow calculated previously~\cite{Petsi2008}. A continuum approach has been applied also in the works~\cite{Widjaja2008,Bhardwaj2009,Son2015}.

Unfortunately, the lack of necessary experimental data impedes development of adequate models. Some experimental data show that transfer of the substances to the edge of the drying droplets is possible only when the Marangoni effect is suppressed~\cite{Hu2006}.
However, other experimental studies demonstrate that the Marangoni effect is the driving force for the formation of a new phase on the edge of a drop.
Moreover, the direction of flow can be opposite to a direction that is predicted by calculations for the pure solvent~\cite{Zaleskiy2004}. Independent experiments confirmed that the flows in pure liquids and in liquids with admixtures go in different directions~\cite{Bardakov2010FD,Chashechkin2010DPh}. In the multi-component liquids of biological origin, the thermo-capillary and soluto-capillary effects can eliminate each other~\cite{Takhistov2002}. Calculations of various research groups have shown that during evaporation of the droplet of a pure liquid there are circular flows caused by the Marangoni stress. The flow is directed along the droplets base to its edge and along its surface towards to the center of the drop~\cite{Girard2006,
Larson2005b,Mollaret2004,Barash2009b,Dunn2008}. At the same time, experiments conducted with biological fluids exhibit opposite flow direction~\cite{Bardakov2010FD,Chashechkin2010DPh,Zaleskiy2004}. It has been suggested that during drying  of the droplets of biological fluids, generally Marangoni flow cannot occur;  the observed circular currents are caused by buoyant convection~\cite{Kistovich2010}.

\section{Crystal growth}\label{sec:crystal}
If salts are present in the droplet, they usually crystallize during drying'
Morphology of the salt crystals is very sensitive to the kind of salt, concentration and type of colloidal particles, as well as the rate of evaporation~\cite{Annarelli2000CE,Annarelli2001,Basu2012,Giri2013,Choudhury2013,Dutta2013COLSUA,DuttaChoudhury2015SM,Roy2015ASS,Yakhno2015,Glibitskiy2015}.
This sensitiveness allows using the morphology of salt crystals as an indicator, e.g. to diagnose different deceases from biological fluids~\cite{Martusevich2007,Martusevich2007BEBM}. At the same time, this sensitivity impedes modelling because a lot of different effects have to be taken into account. In fact, all used models should be treated as semi-empirical. The models often utilize the lattice approach~\cite{Martyushev1997,Martiouchev1998JSP,Martyushev1999TPL,Crivoi2012,DuttaChoudhury2015SM}
and diffusion equation~\cite{Tarasevich2001eng,Dutta2013COLSUA}.
Adequacy of some models~\cite{Martyushev1997,Martiouchev1998JSP,Martyushev1999TPL} has been questioned~\cite{Tarasevich2001eng}. Mainly, dendritic crystal growth can be observed at the final stages of drop desiccation. Both non-equilibrium growth and presence of impurities may produce dendritic shape of crystals~\cite{Langer1980RMPh}; these effects can be reproduced in a simple model~\cite{TarasevichIzvVuzeng}. The phase-field method~\cite{Warren1995} looks extremely promising  for modelling  crystal growth in desiccated colloidal droplets with salt admixtures, but it requires a lot of  additional information, which is difficult to obtain experimentally.
\begin{figure}
  \centering
  \includegraphics[width=\linewidth]{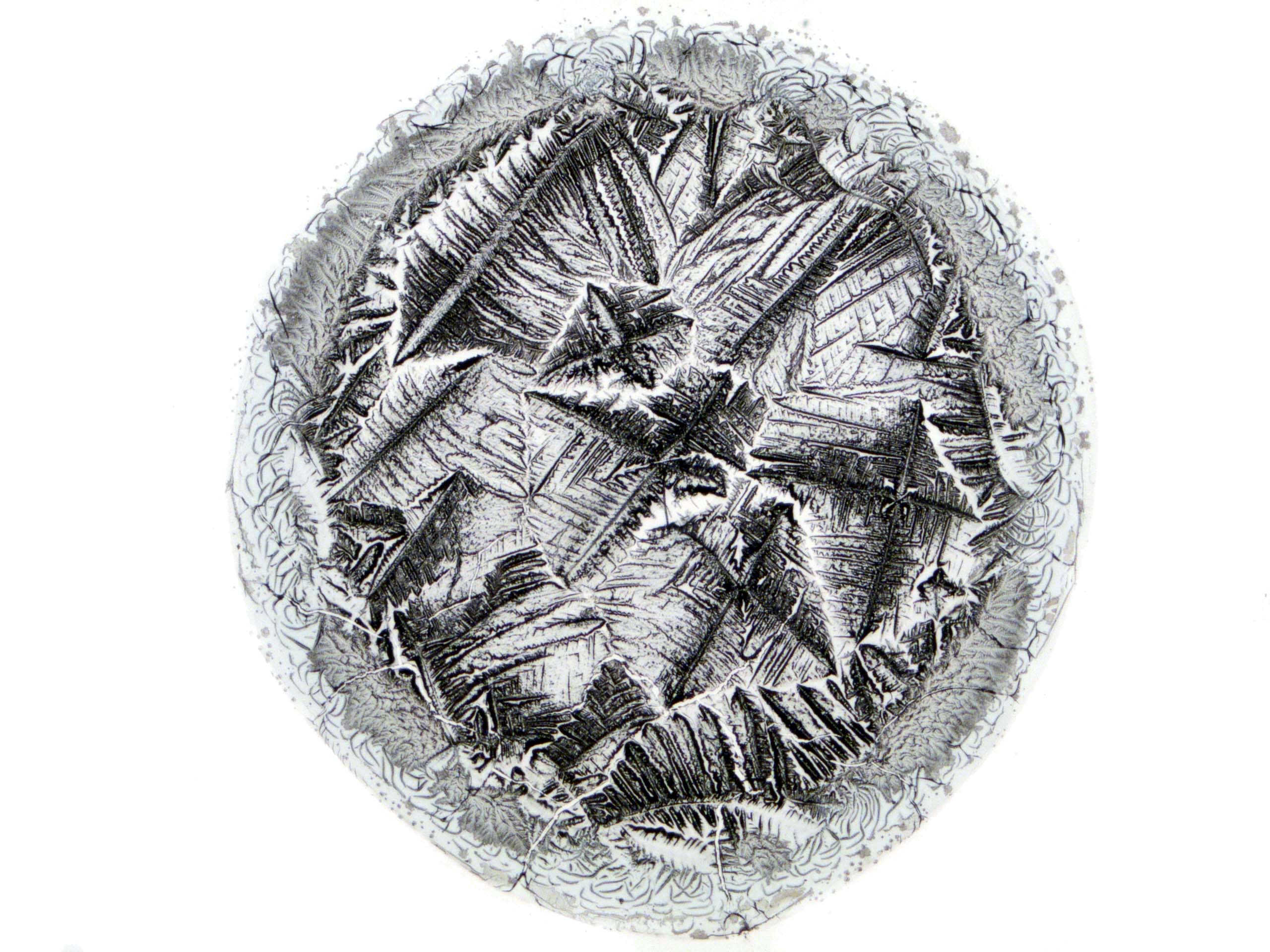}
  \caption{Crystal patterns in a dried sample of albumin + NaCl. \href{http://koltovoi.nethouse.ru/}{Courtesy N.A.~Koltovoi} }\label{fig:koltovoi2}
\end{figure}

\section{Desiccation crack patterns}\label{sec:cracks}
Desiccation crack patterns were intensively investigated both experimentally and theoretically~\cite{Pauchard1999,Annarelli2001,Leung2001Nature,Neda2002,Caddock2002,Jing2012JPhChB,BouZeid2013,Sobac2014CSA,Giorgiutti2014,Ghosh2015Langmuir,Giorgiutti2015SM,Kim2015SR,BouZeid2013,Zhang2013}.
State of the art may be found in the recently published book~\cite{goehring2015desiccation}.
\begin{figure}
  \centering
  \includegraphics[width=\linewidth]{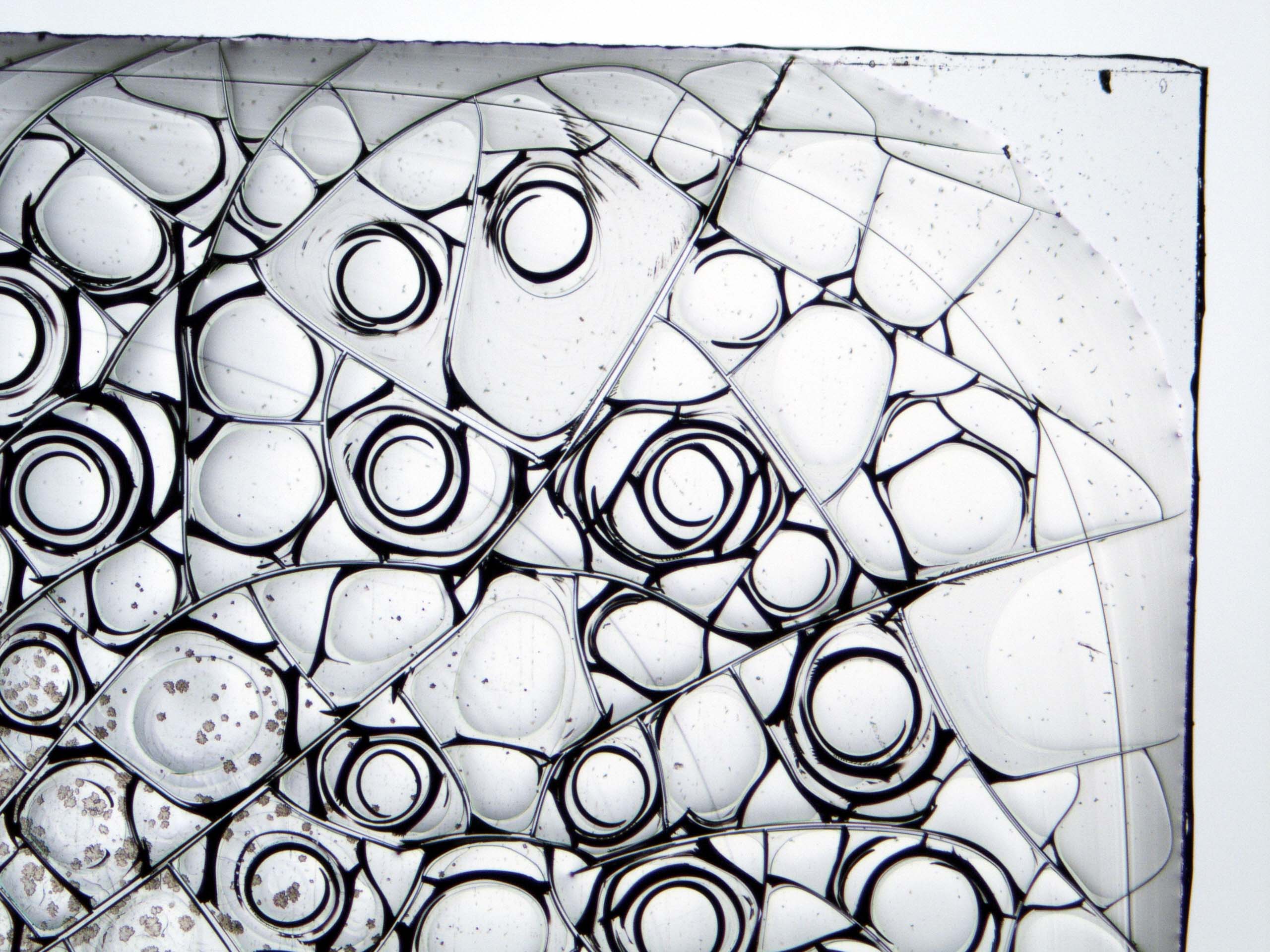}
  \caption{Crack patterns in a dried sample of albumin. \href{http://koltovoi.nethouse.ru/}{Courtesy N.A.~Koltovoi} }\label{fig:koltovoi3}
\end{figure}

While the study of drying droplets is an interesting subject of research that leads to pattern formation, phase segregation and buckling, in some cases the phenomena also result in cracking of the dried drople~\cite{Routh2013,Tsapis2005PRL,Choudhury2013,Annarelli2001}. The simple process of drying a small drop of a multi-component colloidal solution on a flat substrate gives a wealth of
data, which is interesting and potentially useful for practical applications. The crack patterns are used to extract valuable information in various fields such as medical diagnosis, forensics related to crime investigation and material science~\cite{DropletWetting2015}. The final drying pattern and crack nucleation varies with the kinetics of the evaporation rate.  During solvent evaporation, curvature of the solvent-air menisci is responsible for a capillary pressure in the liquid phase. The capillary pressure induces shrinkage of the porous matrix, that is constrained by the adhesion of the deposit to the glass substrate and the evaporation of solvent. As tensile stresses build up, the internal stresses become too great and fractures appear to release mechanical energy. The differences in pattern formation arise due to the competition between the drying process and the adhesion of the matrix on the substrate.

Annarelli et al.~\cite{Annarelli2001} worked on the evaporation, gelling and the cracking behaviour of a deposited drop of protein solution, bovine serum albumin. They observed that the cracks appearing at the gelling edge were regularly-spaced, and were a result of the competition between evaporation induced and relaxation induced evolution. When the crack evolution is only evaporation-induced, the mean crack spacing is proportional to the layer thickness. However in the case of a drop of bovine serum albumin, the evolution of cracks has been described in relation to the change with time of the average shrinkage stress.  In this case the mean crack spacing was observed to be inversely proportional to the deposit thickness. This is unexpected as normally crack spacing increases with thickness.

Brutin and his group worked on the pattern formation of desiccating droplets of human blood from which the coagulation protein had been removed~\cite{DropletWetting2015,Brutin2011FLM}. They  studied the dynamics of the process of evaporation of a blood droplet using a top-view visualization and the drop mass evolution during the drying process. Sobac and Brutin~\cite{Brutin2011FLM} showed that there are two distinct regimes of evaporation during the drying of whole blood. The first regime is driven by convection, diffusion and gelation, while the second regime is only diffusive in nature. A diffusion model of the drying process allows a prediction of the transition between these two regimes of evaporation. Concentration of the solid mass in the drop was important and fracture occurred at a critical mass concentration of solid in a drying drop of blood. They showed that the final crack patterns formed on drying droplets of blood collected from a healthy person, anaemic person and hyperlipidaemic person are quite different. But drawing conclusions for definite diagnosis is not so straightforward as the crack patterns are strongly affected by external
conditions such as the ambient relative humidity and the nature of the substrate.

Brutin et al. conclude that the final drying pattern and crack nucleation varies with the kinetics of the evaporation rate. The transfer of water to air is limited by diffusion and is controlled by the relative humidity in the surrounding air. The drying process of a sessile drop of blood is characterized by an evolution of the solution into a gel saturated with solvent. When the gel is formed, the new porous matrix formed by the aggregation of particles continues to dry by evaporation of the solvent which causes the gel to consolidate. The differences in pattern formation arise due to the competition between the drying process.

Carle and Brutin~\cite{Carle2013lang} studied the influence of surface functional groups and substrate surface energy on the formation of crack patterns and on the dry-out shape in drying a water-based droplet of nano-fluid.
They have also studied
desiccation of blood droplets~\cite{Brutin2012JHT} on different substrates such as glass and glass coated with gold or aluminium.   They measured the rate of heat transfer from the substrate to the  fluid drop. They  show that wettability of the substrate by the fluid is the decisive factor, which can account for the differences in the morphology of the desiccated blood drop on different surfaces, rather than the thermal diffusivity which determines rate of heat transfer from the substrate to the drop. On metallic surfaces where the drop is nearly hemispherical and  a glassy skin forms on the fluid-air interface, there are hardly any cracks. On a glass surface on the other hand, where the drop is more or less flat, an intricate pattern of cracks form.

While there are several works on desiccating droplets done by different groups, there are very few studies on drying droplets in the presence of a perturbation.
The contact angle of a conductive aqueous drop laden with organic or inorganic solutes or ambient oils, changes with the application of Alternating current (AC) voltage during drying. Banpurkar et al.~\cite{Banpurkar2008} studied the above effects in experiments to demonstrate the potential of electrowetting-based tensiometry. Contact angle ($\theta$) decreases with increasing amplitude ($V_{AC}$) of AC voltage following the linear relation of $\cos\theta$ with $V_{AC}$. They applied low frequency AC voltage and obtained interfacial tensions from 5~mJ/m$^{2}$ to 72~mJ/m$^{2}$, in close agreement with the macroscopic tensiometry for drop volumes between 20 and 2000~nL.
Vancauwenberghe et al.~\cite{Vancauwenberghe2013} reviewed the effect of an electric field on a sessile drop. They observed that an external electric field can change the contact angle and shape of a droplet. The electric field also affects the evaporation rate during drying. The contact angle is not always an increasing function of the magnitude of the applied electric field, but may be a decreasing function for some liquid droplets as well.


Khatun et al.~\cite{Khatun2013} investigated desiccation cracks on drying droplets of aqueous Laponite solution, in the presence of a static electric field (DC). The electric field had cylindrical geometry, the peripheral electrode being an aluminium wire bent into a circular form of diameter $\sim 1.8$~cm.  A drop of Laponite gel was deposited inside this wire loop. Another aluminium wire with its tip touching the lower substrate through the centre of the drop, acted as the central electrode. Typical cracks had radial symmetry and were found to emerge always from the positive electrode. With the peripheral electrode positively charged, the final number of cracks $N_{sat}$ appearing on the periphery was measured with the field applied continuously. This was related to the field strength $\Phi$ as
\begin{equation}
N_{sat} = N_s\left(1-\exp\left(-\frac{\Phi}{\Phi_s}\right)\right)
\end{equation}
where $N_s$ is $N_{sat}$ in the limit of a very strong field and $\Phi_s$ is a constant field. The number of cracks at the centre, when the central electrode is positive, was  found to be 3 in most cases and rarely 4.

The time of appearance of the first crack $t_{cr}$ after deposition of the drop, was also found to be a function of the field strength.
\begin{equation}
t_{cr} = t_0 \exp\left(-\frac{\Phi}{\Phi_0}\right).
\end{equation}
Here $t_0$ is the time of appearance of the first crack in the absence of any electric field. $\Phi_0 $ is a constant field such that when   $\Phi = \Phi_0$,  $t_{cr} $ falls to $t_0/e$.

In another set of experiments, the authors applied the field for a very short time $\tau$ and switched it off before
any crack appeared.  In this case the final cracks showed the same pattern as when the field was always on. However in this case, the time of appearance of the first crack was delayed to $t_a(\Phi,\tau)$ and the final number of cracks reduced to $n_f(\Phi,\tau)$.
These quantities were found to obey simple empirical rules on appropriate transformations of the variables. The rules are
\begin{equation}
\frac{(t_a-t_{cr})\Phi_0}{\Phi t_0} \propto \frac{\tau}{t_{cr}} \label{eq:cr-time}
\end{equation}
and
\begin{equation}
\frac{(N_{sat}-n_f)\Phi_s}{\Phi N_s} \propto \frac{\tau}{t_{sat}} \label{eq:cr-num}
\end{equation}
These relations quantify the strength and duration of the memory of electric field exposure  retained by the sample. Following equations \eqref{eq:cr-time} and \eqref{eq:cr-num}, the data for $t_{cr}$ and $n_f$ collapsed on a master graph for different applied voltages. These results are interesting because the collapse of data sets with different applied voltage $\Phi$ indicates that the transformed crack appearance times, or rather the extra time needed for crack appearance when the external field is switched off before $t_{cr}$,  increases in proportion to $\Phi$ and is a simple function of the non-dimensionalised field exposure time. The same is true of the number of cracks, with a reversal in sign, since the number of cracks for $\tau = t_{cr}$ is more than the number when $\tau < t_{cr}$. However, a theoretical interpretation of the scaling relations is not yet understood.

The authors further calculated the energy dissipated $U_{dis}$ in the sample when the field is on and upto the appearance of the first crack, according to
\begin{equation}
 U_{dis} = \int_0^{t_c} \Phi I(t) \mathrm{d}t.
 \end{equation}
Here $I(t)$ is the current through the system at time $t$. $U_{dis}$ was found to be more or less constant for all applied voltages.

Scanning electron microscopy (SEM) was done on different regions of the droplet. The images showed that samples without field exposure or with low field exposure, have more micro-cracks compared
to samples exposed to high fields. Field exposure induces large cracks which release stress, so regions between these large cracks are relatively flawless. These studies
indicate that direct and alternating electric fields both have a strong effect on the formation of desiccation cracks and may presumably be used for intentionally promoting or suppressing crack growth.


\section{Conclusion}\label{sec:concl}
There obviously remains much more work to be done in this interesting and useful area of research.
Regarding directions of research, some tasks should be especially emphasized
\begin{itemize}
\item Obtaining new experimental data, critically needed for the design and development of adequate models.
\item The development of 3D models describing the redistribution of the components, the movement of the phase front and the evolution of the profile of the drying colloidal droplets with salt admixtures. In these systems, phase transition from sol to gel is concentration driven. The thermal phase transition from liquid to vapor also takes place in this system. This phase transition leads to a movement of the liquid-vapor phase boundary, i.e. the droplet volume decreases and droplet profile changes.
\item Considering additional effects that may be crucial to understanding the processes of pattern formation, but have not yet been included in the model, e.g. variations of the viscosity of a colloid with time and concentration of salts, changes of the vapor flux above the free surface of the droplets when the phase boundary (sol-gel) is moving.
\item Analyzing the final pattern through tools such as fractal and multifractal characterization.
\end{itemize}

\section*{Acknowledgements}
The reported study was funded by the Ministry of Education and Science of the Russian Federation according to the research  Project No.~643 (Yu.T). S. Tarafdar thanks DST, Govt. of India for the grant of a research project (SR/S2/CMP-127/2012).

\bibliography{drops2016}
\end{document}